\documentclass[11pt]{article}

\usepackage{amsmath}
\usepackage{graphicx}
\usepackage{amsfonts}
\usepackage{amssymb}
\usepackage{epsfig}
\usepackage{color}
\usepackage{psfrag}
\usepackage{epstopdf}

\usepackage{caption}
\usepackage{subcaption}

\newcommand{\EQ}{\begin{equation}}
\newcommand{\EN}{\end{equation}}
\newcommand{\be}{\begin{equation}}
\newcommand{\ee}{\end{equation}}
\newcommand{\bea}{\begin{eqnarray}}
\newcommand{\eea}{\end{eqnarray}}

\begin{document} \setcounter{page}{0}
\newpage
\setcounter{page}{0}
\renewcommand{\thefootnote}{\arabic{footnote}}
\newpage
\begin{titlepage}
\begin{flushright}
\end{flushright}
\vspace{0.5cm}
\begin{center}
{\large {\bf Exact results for spin glass criticality}}\\
\vspace{1.8cm}
{\large Gesualdo Delfino}\\
\vspace{0.5cm}
{\em SISSA -- Via Bonomea 265, 34136 Trieste, Italy}\\
{\em INFN sezione di Trieste, 34100 Trieste, Italy}\\
\end{center}
\vspace{1.2cm}

\renewcommand{\thefootnote}{\arabic{footnote}}
\setcounter{footnote}{0}

\begin{abstract}
\noindent
In recent years scale invariant scattering theory provided the first exact access to the magnetic critical properties of two-dimensional statistical systems with quenched disorder. We show how the theory extends to the overlap variables entering the characterization of spin glass properties. The resulting exact fixed point equations yield both the magnetic and, for the first time, the spin glass renormalization group fixed points. For the case of the random bond Ising model, on which we focus, the spin glass subspace of solutions is found to contain a line of fixed points. We discuss the implications of the results for Ising spin glass criticality and compare with the available numerical results.
\end{abstract}
\end{titlepage}

\newpage
\tableofcontents

\section{Introduction}
Random magnets are expected to possess a spin glass phase for strong enough frustration at low temperatures \cite{EA,BY,KR}. While the mean field solution is known \cite{Parisi} for the basic model with infinite range interaction \cite{SK}, analytically accessing the spin glass phase for the case of short range interactions in two and three dimensions proved to be extremely difficult and the problem essentially remained the realm of numerical studies. The accuracy of the latter, on the other hand, is limited by the difficulties related to the need of reaching large system sizes \cite{BY,KR}.

In recent years it has been shown that the replica method normally used to deal with quenched disorder can be implemented in an exact way when working directly at criticality in two dimensions \cite{random}. The key observation is that in this case conformal symmetry \cite{DfMS} can be exploited in a simple way within the scattering framework of the underlying field theory, thus providing the first exact access to random criticality. In the language of the renormalization group \cite{Cardy_book}, the method yields the fixed points whose basins of attraction determine the critical behavior in the different regions of the phase diagram. So far the scattering method has been used for the study of the ``magnetic" random fixed points \cite{random,DT2,DL_ON1,DL_ON2,DL_softening,random_line} (\cite{colloquium} for a review), namely those fixed points located along the boundary of the ferromagnetic phase. In the present paper we show how the method extends to the spin glass renormalization group fixed points, namely the fixed points ruling the critical exponents inside the spin glass regime. This is achieved introducing in the scattering framework the ``overlap" variables entering the characterization of the spin glass properties. In a nontrivial way, the resulting exact fixed point equations make explicit the specific nature of the overlap variables, which account for correlations among independent copies of the system with the same disorder distribution. Remarkably, for the basic example of the random bond Ising model that we analyze, the spin glass subspace of solutions of the exact equations turns out to contain a line of fixed points. We then discuss the implications of the results from the point of view of universality and in comparison with the available numerical results for the spin glass regime.

The paper is organized as follows. After recalling in the next section some main features of the two-dimensional random bond Ising model, we derive the exact renormalization group fixed point equations in section~\ref{eqs}. The solutions are discussed in section~\ref{solutions} before collecting some final remarks in the final section.

\section{Magnetic and spin glass criticality}
\begin{figure}[t]
\centering
\includegraphics[width=6.5cm]{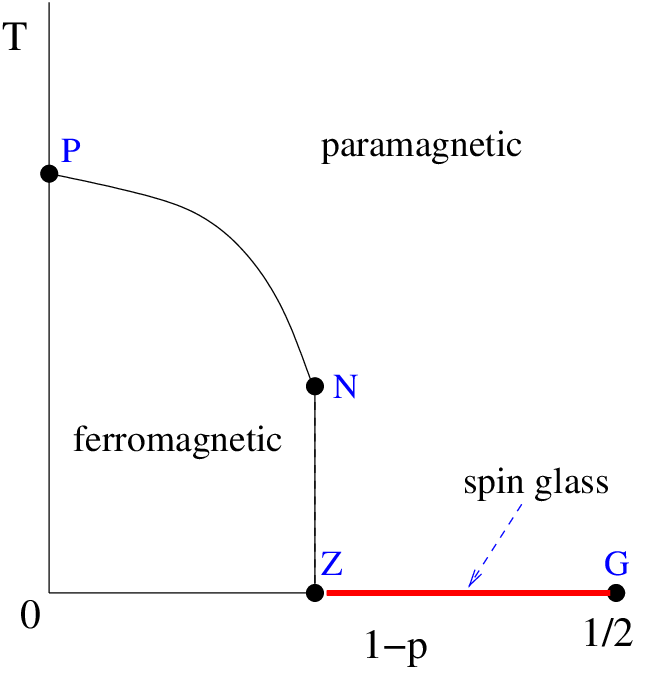}
\caption{Phase diagram of the $\pm J$ random bond Ising model in two dimensions. The renormalization group fixed points correspond to the critical point of the pure model P, the Nishimori point N, the zero temperature magnetic fixed point Z, and the spin glass fixed point G.}
\label{pd}
\end{figure}

The random bond Ising model \cite{EA} is defined by the Hamiltonian
\EQ
H=-\sum_{\langle x,y\rangle}J_{xy}\,\sigma(x)\sigma(y)\,,
\label{Ising}
\EN
where $\sigma(x)=\pm 1$ denotes the spin variable at site $x$ of a regular lattice, $\sum_{\langle x,y\rangle}$ is the sum over nearest neighbors, and quenched disorder is introduced through random couplings $J_{xy}$ drawn from a probability distribution $P(J_{xy})$. A common choice is the bimodal distribution
\EQ
P(J_{xy})=p\,\delta(J_{xy}-1)+(1-p)\,\delta(J_{xy}+1)\,,
\label{bimodal}
\EN
where $1-p$ is the fraction of antiferromagnetic bonds. The ordered phase of the pure ferromagnet ($p=1$) persists as the disorder strength $1-p$ is increased, until the transition temperature vanishes. The transition between this ferromagnetic phase and the paramagnetic phase (Figure~\ref{pd}) is continuous, namely the correlation length $\xi_M$ defined by\footnote{We denote by $\langle\cdots\rangle$ the thermal average and by $[\cdots]$ the average over the random variables $J_{xy}$.}
\EQ
[\langle\sigma(x)\sigma(y)\rangle]_\textrm{conn}\sim e^{-|x-y|/\xi_M}\,,\hspace{1cm}|x-y|\to\infty
\label{xi_m}
\EN
diverges along the phase boundary. We refer to the divergence of $\xi_M$ as {\it magnetic criticality}. 
In the two-dimensional case of interest in the present paper, magnetic criticality is ruled by the three renormalization group fixed points present on the ferromagnetic-paramagnetic phase boundary. The fixed point P of the pure model is analytically known to be stable under the addition of disorder \cite{DD_81,DD_83,Shalaev,Ludwig}; numerical studies \cite{McMillan,MC,PHP,HPtPV} show that the zero temperature fixed point Z is stable in the temperature direction, while the Nishimori point N \cite{Nishimori,Nishimori_book,LdH1,LdH2} is unstable in all directions. Recently, the exact correlation length and crossover critical exponents at the Nishimori point have been determined exactly (also in three dimensions) \cite{GD_Nishimori} starting from properties of the Nishimori line \cite{Nishimori,Nishimori_book}. At a given fixed point, the behavior
\EQ
[\langle\sigma(x)\sigma(y)\rangle]\sim |x-y|^{-2X_\sigma}
\EN
defines the spin scaling dimension $X_\sigma$.

In general, spin glass behavior is expected to arise as an effect of randomness and frustration at low enough temperatures inside the region of zero magnetization. It is investigated introducing the overlap variable \cite{EA,BY}
\EQ
q^{a,b}(x)\equiv\sigma^{(a)}(x)\sigma^{(b)}(x)\,,\hspace{1cm}a\neq b\,,
\label{overlap}
\EN
where $a$ and $b$ label independent copies\footnote{These copies are also called {\it real} replicas, in order to distinguish them from the $n\to 0$ auxiliary replicas entering the replica method, see below.} of the system with the same disorder realization $\{J_{ij}\}$. The divergence of the correlation length $\xi_{O}$ defined by
\EQ
[\langle q^{a,b}(x)q^{a,b}(y)\rangle]_\textrm{conn}=[\langle\sigma(x)\sigma(y)\rangle^2]_\textrm{conn}\sim e^{-|x-y|/\xi_O}\,,\hspace{1cm}|x-y|\to\infty
\label{xi_o}
\EN
corresponds to {\it overlap criticality}, while the fixed point behavior
\EQ
[\langle q^{a,b}(x)q^{a,b}(y)\rangle]\sim |x-y|^{-2X_q}
\EN
defines the overlap scaling dimension $X_q$. {\it Spin glass criticality} amounts to overlap criticality outside the ferromagnetic region and its boundary. Numerical consensus developed since \cite{McMillan} (see also \cite{Nishimori_T=0} for analytical arguments) on the fact that in two dimensions there is no finite temperature transition to a spin glass phase and that spin glass criticality occurs only at $T=0$. The overlap correlation length is observed to diverge as $\xi_O\sim T^{-\nu_{SG}}$ and the value of $\nu_{SG}$ is expected to be ruled by a zero temperature spin glass fixed point G (see Figure~\ref{pd}). On bipartite lattices, for which the phase diagram is symmetric under $p\to 1-p$, G is expected to fall at $p=1/2$; many numerical studies of spin glass properties focus on this value.

\section{Exact fixed point equations}
\label{eqs}
The discussion of the previous section shows that the study of overlap properties amounts to consider the Hamiltonian
\EQ
H_N=-\sum_{\langle x,y\rangle}J_{xy}\sum_{a=1}^N\sigma^{(a)}(x)\sigma^{(a)}(y)
\label{lattice}
\EN
of $N$ decoupled Ising copies with the same disorder realization. For $N=1$ one recovers the Hamiltonian (\ref{Ising}) which is sufficient for the study of the magnetic properties. For $N\neq 1$ different copies are correlated by the disorder average. Quenched disorder corresponds to the fact that the average over the random variables $J_{xy}$ is taken on the free energy $-\ln Z$, where 
\EQ
Z=\sum_{\{\sigma^{(1)}(x)\},\cdots,\{\sigma^{(N)}(x)\}}e^{-H_N/T}
\label{Z}
\EN
is the partition function with an assigned disorder configuration. Theoretically, it is convenient to exploit the identity 
\EQ
\ln Z=\lim_{n\to 0}\frac{Z^n-1}{n}\,,
\EN
through which the average over disorder has the effect of coupling $n\to 0$ replicas of the system with Hamiltonian (\ref{lattice}). 

It was shown in \cite{random} (see \cite{colloquium} for a review) that in two dimensions the replica method can be implemented in an exact way at renormalization group fixed points, which in the continuum correspond to conformal field theories. One uses the scattering framework of \cite{paraf}, in which one of the two spatial dimensions plays the role of imaginary time, and exploits the fact that infinite-dimensional conformal symmetry in two dimensions \cite{DfMS} yields infinitely many quantities conserved in scattering processes. This forces initial and final states to be kinematically identical (complete elasticity), while scale and relativistic invariance make scattering amplitudes energy independent. These simplifications of the scattering problem are strong enough to make it exactly solvable, a circumstance that has been exploited to progress with longstanding problems of two-dimensional criticality, both with \cite{random,DT2,DL_ON1,DL_ON2,DL_softening,random_line} and without disorder \cite{DT1,ising_vector,DDL_nematic,DLD_RPN,DLD_CPN,potts_qr,RPN_universality}.

\begin{figure}[t]
\centering
\includegraphics[width=14cm]{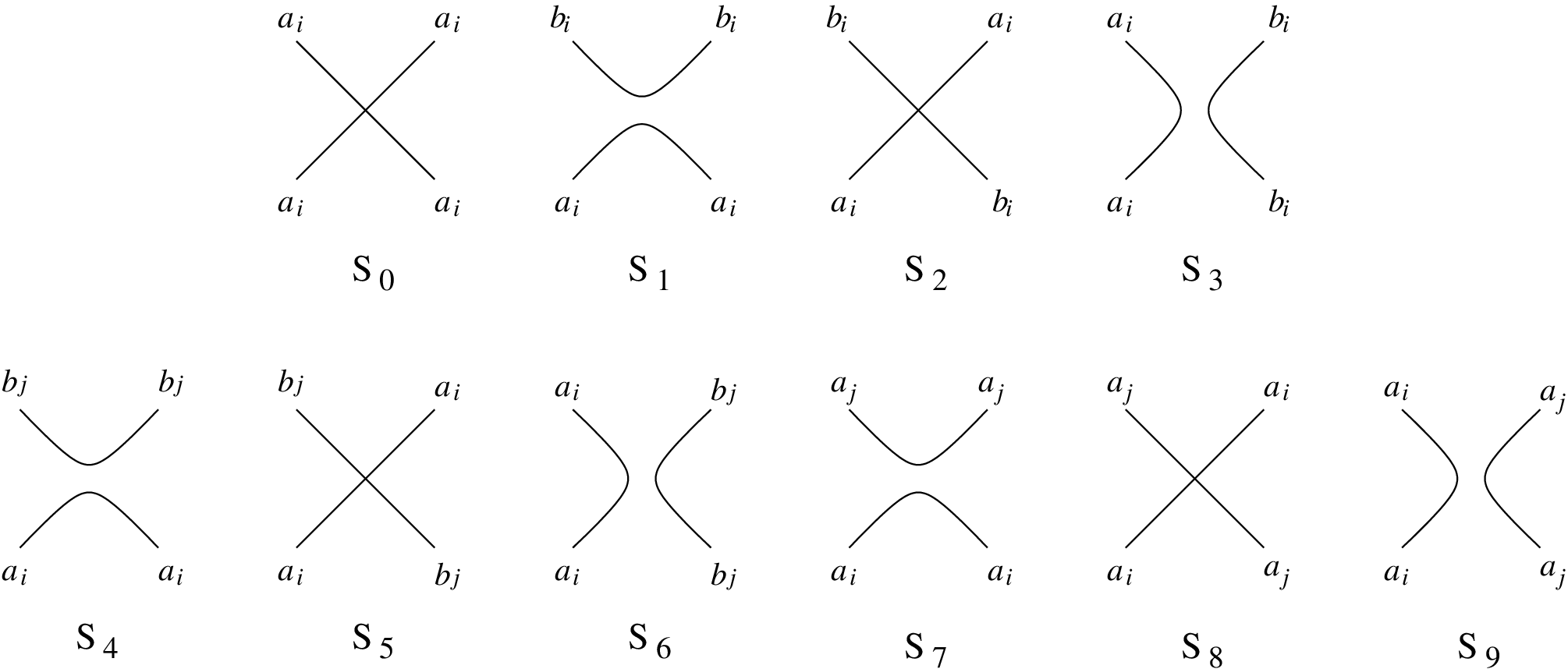}
\caption{Magnetic and overlap scattering amplitudes for the random bond Ising model. $a_i$ labels a particle excitation in the $i$-th replica of the $a$-th copy ($a\neq b$, $i\neq j$). Overlap amplitudes involve different copies. Time runs upwards.}
\label{amplitudes}
\end{figure}

In this framework, the scattering particles correspond to the fundamental collective excitation modes of the system, and the symmetry representation they carry characterizes the model under consideration. In the present case, $a=1,2,\ldots,N$ labels the particle corresponding to an elementary excitation in the $a$-th Ising copy; such a particle is odd under spin reversal $\sigma^{(a)}\to-\sigma^{(a)}$, which is a symmetry of the Hamiltonian (\ref{lattice}). In addition, an index $i=1,2,\ldots,n$ labels the replicas associated to the disorder average. We then denote by $a_i$ a particle in the $i$-th replica of the $a$-th Ising copy. The elastic scattering amplitudes allowed by spin reversal symmetry and by permutational symmetry (of the $N$ copies as well as of the $n$ replicas) are shown in Figure~\ref{amplitudes}.

As for any scattering theory \cite{ELOP}, the amplitudes satisfy crossing and unitarity equations, which at two-dimensional fixed points take a particularly simple form as a consequence of complete elasticity and energy independence. With the notation $S_{\alpha\beta}^{\gamma\delta}={}_\alpha^\delta\times_\beta^\gamma$ for a generic amplitude, they read \cite{colloquium,paraf}
\begin{equation}
S_{\alpha\beta}^{\gamma\delta}=[S_{\alpha\delta}^{\gamma\beta}]^*\,,
\label{cross}
\end{equation} 
\begin{equation}
\sum_{\epsilon,\phi} S_{\alpha\beta}^{\epsilon\phi}[S_{\epsilon\phi}^{\gamma\delta}]^*=\delta_{\alpha\gamma}\delta_{\beta\delta}\,,
\label{unitarity}
\end{equation}
respectively. The amplitudes are unchanged under time reversal ($S_{\alpha\beta}^{\gamma\delta}=S^{\alpha\beta}_{\gamma\delta}$) and spatial reflection ($S_{\alpha\beta}^{\gamma\delta}=S_{\beta\alpha}^{\delta\gamma}$). When specialized to the amplitudes of Figure~\ref{amplitudes}, the crossing equations (\ref{cross}) translate into
\begin{align}
S_k&=S_k^*\,,\hspace{3.4cm}k=0,2,5,8
\label{crossing1}\\
S_k&=S_{k+2}^*\equiv X_k+iY_k\,,\hspace{1cm}k=1,4,7\,,
\label{crossing2}
\end{align}
where we introduced the real quantities $X_k$ and $Y_k$. It follows that the unitarity equations (\ref{unitarity}) can be written as
\begin{align}
&S_0^2+(N-1)[X_1^2+Y_1^2+(n-1)(X_4^2+Y_4^2)]+(n-1)(X_7^2+Y_7^2)=1\,,\label{uni1}\\
&2S_0 X_1+(N-2)[X_1^2+Y_1^2+(n-1)(X_4^2+Y_4^2)]+2(n-1)(X_4 X_7+Y_4 Y_7)=0\,,\label{uni2}\\
&2 S_0 X_4+2(X_1X_7+Y_1Y_7)+(N-2)[2(X_1X_4+Y_1Y_4)+(n-2)(X_4^2+Y_4^2)]\nonumber\\
&+2(n-2)(X_4X_7+Y_4Y_7)=0\,,\label{uni3}\\
&2 S_0X_7+(n-2)(X_7^2+Y_7^2)+(N-1)[2(X_1X_4+Y_1Y_4)+(n-2)(X_4^2+Y_4^2)]=0,\\
&X_k S_{k+1}=0\,,\hspace{2.4cm}k=1,4,7\,,\label{uni9}\\
&X_k^2+Y_k^2+S_{k+1}^2=1\,,\hspace{1cm}k=1,4,7\,.
\label{uni10}
\end{align}
These equations contain $N$ and $n$ as parameters that do not need to be integers. In particular, the limit $n\to 0$ yielding quenched disorder can be taken straightforwardly.

The derivation of Eqs.~(\ref{uni1})-(\ref{uni10}) implies that their solutions are the renormalization group fixed points of a system of $Nn$ Ising models constrained by invariance under spin reversals $\sigma_i^{(a)}\to -\sigma_i^{(a)}$ and permutations of the $N$ copies and of the $n$ replicas. These symmetries, however, allow for the addition to the Hamiltonian (\ref{lattice}) of terms coupling the different copies\footnote{This is why Eqs.~(\ref{uni1})-(\ref{uni10}) also arise in the study of the disordered $N$-color Ashkin-Teller model \cite{random_line}.}, so that the space of solutions of (\ref{uni1})-(\ref{uni10}) still includes fixed points which are not of interest for the problem we study in this paper. In our present case the Ising copies are correlated only by the disorder average, and their number $N$ is not a physical parameter of the problem. This requires that, for $n=0$, the space of solutions of (\ref{uni1})-(\ref{uni10}) contains a $N$-independent subspace corresponding to the fixed points of our interest. Remarkably, it is immediately seen that, at $n=0$, the $N$-dependent terms in (\ref{uni1})-(\ref{uni10}) cancel under the condition
\EQ
S_4=S_1\,.
\label{s4}
\EN
The resulting equations read
\begin{align}
&S_0^2-(X_7^2+Y_7^2)=1\,,\label{u1}\\
&S_0 X_1-(X_1 X_7+Y_1 Y_7)=0\,,\label{u2}\\
&S_0X_7-(X_7^2+Y_7^2)=0\,,\\
&X_k S_{k+1}=0\,,\hspace{2.4cm}k=1,7\,,\label{u4}\\
&X_k^2+Y_k^2+S_{k+1}^2=1\,,\hspace{1cm}k=1,7\,,
\label{u5}
\end{align}
where we took into account that (\ref{uni2}) and (\ref{uni3}) lead to the same equation; we also have 
\EQ
S_5=\pm S_2\,. 
\label{s5}
\EN

\section{Solutions and discussion}
\label{solutions}

\begin{table}
\begin{center}
\begin{tabular}{c|c|c|c|c|c}
\hline 
Solution & $S_0$ & $X_7$ & $Y_7$ & $X_1$ & $Y_1$\\ 
\hline \hline
$\text{A}_{\pm}$ & $\pm 1$ & $0$ & $0$ & $0$ & $[-1,1]$ \\ 
$\text{B}_{\pm}$ & $\pm\sqrt{2}$ & $\pm\frac{1}{\sqrt{2}}$ & $(\pm)\frac{1}{\sqrt{2}}$ & $0$ & $0$ \\ 
$\text{C}_{\mp}$ & $\sqrt{2}$ & $\frac{1}{\sqrt{2}}$ & $\mp\frac{1}{\sqrt{2}}$ & $\mp Y_1$ & $(\pm)\frac{1}{\sqrt{2}}$ \\
$\text{D}_{\pm}$ & $-\sqrt{2}$ & $-\frac{1}{\sqrt{2}}$ & $\pm\frac{1}{\sqrt{2}}$ & $\mp Y_1$ & $(\pm)\frac{1}{\sqrt{2}}$ \\
[0.7em] 
\hline 
\end{tabular} 
\caption{Solutions of Eqs.~(\ref{u1})-(\ref{u5}). They yield the magnetic and spin glass renormalization group fixed points of the random bond Ising model. Signs in parenthesis are both allowed.
}
\label{sol}
\end{center}
\end{table}

The solutions of Eqs.~(\ref{u1})-(\ref{u5}) are listed in Table~\ref{sol}. As discussed in the previous section, they yield the magnetic and spin glass renormalization group fixed points of the random bond Ising model (Figure~\ref{pd}), and our next task is to identify which fixed point corresponds to which solution. For this purpose, we must first of all observe that the form of the crossing and unitarity equations (\ref{cross}) and (\ref{unitarity}) is such that, given a solution, another solution is obtained reversing the sign of all the scattering amplitudes. It follows that the space of solutions of the exact fixed point equations provided by scattering theory in two dimensions is always made of pairs of solutions related by such ``sign reversal". In some cases only one member of a pair is of physical interest, the other being automatically generated by the form of the equations. In our present case, the pairs are $A_+$ and $A_-$, $B_+$ and $B_-$, as well as $C$ and $D$. 

A second observation is that the solutions of Table~\ref{sol} exhibit also another type of pairing, this time specific to the problem we are studying. Indeed, given the values of $S_0$ and $S_7=X_7+iY_7$, there is always a solution with $S_1=X_1+iY_1=0$ and one with $S_1\neq 0$. These pairs are $A_\pm|_{Y_1=0}$ and $A_\pm|_{Y_1\neq 0}$, $B_+$ and $C$, as well as $B_-$ and $D$. The physical meaning of this result is understood once we recall that $S_0$ and $S_7$ characterize the magnetic sector\footnote{Recall that $S_8$ and $S_9$ are related to $S_7$ through (\ref{crossing2}) and (\ref{u5}).}, namely the amplitudes which are well defined already for $N=1$ and describe the interaction within a single copy. On the other hand, $S_1$ characterizes the overlap sector\footnote{Recall that $S_2,\dots,S_6$ are related to $S_1$ through (\ref{crossing2}), (\ref{s4}), (\ref{u5}) and (\ref{s5}).} containing the amplitudes which involve different copies. Hence, the solutions with $S_1=0$ (no overlap) correspond to the fact that the magnetic fixed points can be studied with no reference to overlaps ($N=1$). On the other hand, the partner solutions with $S_0$ and $S_7$ unchanged and $S_1\neq 0$ reflect the fact that the overlap properties do not affect the magnetic ones. 

With this understanding, we begin with the identification of the solutions corresponding to the magnetic fixed points P, N and Z of Figure~\ref{pd}. The fact that the two-dimensional Ising ferromagnet without disorder is a theory of free fermions \cite{DfMS,Onsager} implies that P corresponds to\footnote{Scattering on a line is related to position exchange and statistics. A transmission amplitude $-1$ corresponds to a free fermion.} $A_-|_{Y_1=0}$. In absence of disorder $[\langle q^{a,b}(x)q^{a,b}(y)\rangle]=\langle\sigma(x)\sigma(y)\rangle^2$, so that at P there is also trivial overlap criticality (uncorrelated copies) with $X_q=2X_\sigma$. 

Turning to the fixed points N and Z, we can exploit the fact that the scattering theory was used to study the magnetic sector of the random bond $q$-state Potts model \cite{random,colloquium}, which yields the Ising model for $q=2$. The $q$-dependence of the different solutions allowed to identify the magnetic sector of the Ising fixed points N and Z with our present solution $B_-$, the two points being distinguished by the sign of $Y_7$. On the other hand, the identity $[\langle q^{a,b}(x)q^{a,b}(y)\rangle]=[\langle\sigma(x)\sigma(y)\rangle]$, which is known \cite{Nishimori,Nishimori_book} to hold at the Nishimori point N, implies that at this point there is also overlap criticality with $X_q=X_\sigma$. This simultaneous magnetic and overlap criticality corresponds to the partner solution of $B_-$ with $S_1\neq 0$, namely to solution $D$. The coincidence at point N of the spin and overlap correlation functions indicates that the critical behaviors in the two sectors coincide at this point, namely that $S_1=S_7$. It is then no accident that solution $D$ indeed allows for this identity. Overlap criticality is expected in the whole region from Z to G. Hence, simultaneous magnetic and overlap criticality at the fixed point Z corresponds again to solution $D$, the difference with N being made by sign choices. 

Finally, we should identify the solution corresponding to the spin glass fixed point G. The absence of magnetic criticality at G excludes the solutions $A_-$, $B_-$ and $D$ which, as we saw, account for the fixed points P, N and Z. In addition, solution $B_+$ corresponds to uncorrelated copies ($S_1=0$). We are then left with the solutions $A_+$ and $C$. They correspond to different scenarios which are better understood after recalling some general facts about universality. 

Universality is the property of classes of statistical systems with short range interactions and the same internal symmetry of exhibiting the same critical exponents despite microscopic differences such as a different lattice structure (e.g., in two dimensions, square vs triangular). For ferromagnets, universality arises because the tendency of the spins to point in the same direction makes the number of neighbors irrelevant in the renormalization group sense. For antiferromagnets, on the contrary, the number of neighbors is essential in establishing to which extent the tendency of nearby spins to point in different directions can be fulfilled. As a consequence, antiferromagnets with the same Hamiltonian but different lattice structure can renormalize on different fixed points. Also in this respect, the ability of the scattering theory in two dimensions to give at once the different fixed points allowed by a given symmetry is remarkable and was illustrated in \cite{colloquium,DT1} for the $q$-state Potts model. Here we simply recall that $q=3$ allows for a line of fixed points\footnote{This is the line of fixed points of the Gaussian model, which also makes possible a number of other nontrivial phenomena in two-dimensional criticality, such as the BKT transition \cite{Berezinskii,KT}, the continuously varying exponents of the Ashkin-Teller model \cite{Baxter,KB}, and the Luttinger liquid behavior of electrons in ($1+1$) dimensions \cite{Tomonaga,Luttinger,ML}.}, and that it was predicted in \cite{fpu} that three-state Potts antiferromagnets on different bipartite lattices can renormalize on different points of this line; it was numerically shown in \cite{q3lattice} that this is indeed the case. 

Clearly, frustrated random magnets, in which ferromagnetic and antiferromagnetic bonds are both allowed, may in principle exhibit nonuniversal critical behavior depending on the lattice structure. Moreover, changing the disorder distribution $P(J_{xy})$ is an additional potential source of nonuniversality. When weak disorder is relevant in the renormalization group sense -- for Ising this is the case in three dimensions but not in two -- before the Nishimori point there is an additional fixed point which can be perturbatively shown \cite{Cardy_book} to be universal in the class of the randomly dilute ferromagnet. In our present two-dimensional case, universality for the magnetic fixed points is consistent with our results as well as with the conclusions of numerical studies which compared different lattices and disorder distributions (see \cite{PHP,deQueiroz}). The spin glass fixed point G, on the other hand, has strongest disorder and maximizes the chances of nonuniversal behavior. There are two main scenarios depending on whether G corresponds to solution $C$ or to solution $A_+$. Solution $C$ leads to a single fixed point, or at most a few if different sign choices are physically relevant. This in turn amounts to a single universality class or to a small discrete set of universality classes. Solution $A_+$, instead, possesses $Y_1$ as a free parameter and corresponds to a {\it line of fixed points}. Different lattices and/or different disorder distributions may then renormalize on arbitrarily close points along this line. 

At present we do not see a way to theoretically single out one of the two scenarios. On the side of numerical simulations, on the other hand, the issue of universality at the fixed point G is still debated \cite{KLC,Hartmann,PtPV,Fernandez1,LC,KW}. The most accurate measurement to date is that of \cite{KW} for the domain wall fractal dimension on the square lattice, which gave $1.27319(9)$ for the Gaussian disorder distribution and $1.279(2)$ for the bimodal distribution. These authors considered the two results as ``marginally consistent" with each other and concluded that they leave open the question of universality at the spin glass fixed point G. Our results now allow for the possibility that the two distributions renormalize on close but distinct points on the line $A_+$. Additional interesting insight comes from the numerical conclusion of \cite{Fernandez2} that the square lattice model with the disorder (\ref{bimodal}) and $p=1/2$ behaves ``almost but not quite" as a free boson. Since solution $A_+|_{Y_1=0}$ is a free boson (pure transmission with amplitude $+1$), $A_+|_{Y_1\ll 1}$ is almost a free boson. Solution $C$, instead, is quite far from free bosonic behavior.

\section{Conclusion}
Scale invariant scattering theory has been used in the last years to obtain the first exact results for renormalization group fixed points of two-dimensional statistical systems with quenched disorder. So far, the results had concerned the magnetic fixed points, namely the fixed points located on the ferromagnetic-paramagnetic phase boundary. In this paper we showed how the theory can be extended to include the overlap variables which characterize the spin glass behavior. The overlaps account for the correlations among independent copies (real replicas) of the system with the same disorder realization. Quite remarkably, the theory explicitly shows how the initial set of fixed point equations produced by the symmetry constraints simplifies under the requirement that the different copies are correlated only by the disorder average. The space of solutions of this final set of exact equations contains both the magnetic and the spin glass renormalization group fixed points. The magnetic fixed points coincide with those obtained within the theory which does not detect the overlaps \cite{random,colloquium}, as they should. The exact access to the spin glass sector, on the other hand, is obtained here for the first time.

For the random bond Ising model, on which we focused, we showed that the subspace of spin glass solutions contains an exact line of fixed points. This is a remarkable finding in view of the fact that it implies the rarely fulfilled renormalization group condition of ``true marginality". We discussed the scenarios allowed by these results for Ising spin glass critical behavior, for which strong frustration makes the issue of universality particularly nontrivial. We then observed that the available numerical results are consistent with the solution corresponding to the line of fixed points.

\end{document}